\documentclass[conference]{IEEEtran}
\usepackage{booktabs}
\usepackage{tikz}
\usepackage{url}
\usepackage{pdfpages}
\usepackage{afterpage}
\usepackage{color, colortbl}
\usepackage{graphicx}
\usepackage{amsmath}
\usepackage{import}
\usepackage{pifont}
\usepackage{multirow}
\usepackage{subfigure}
\usepackage{footmisc}
\usepackage{algorithm}
\usepackage{lipsum}
\usepackage[
separate-uncertainty = true,
multi-part-units = repeat
]{siunitx}
\usepackage[noend]{algpseudocode}

\def\BState{\State\hskip-\ALG@thistlm}

\definecolor{Gray}{gray}{0.9}
\definecolor{LightCyan}{rgb}{0.88,1,1}
\pdfoutput=1
\graphicspath {{figures/}}

\makeatletter
\DeclareRobustCommand*\textsubscript[1]{%
  \@textsubscript{\selectfont#1}}
\def\@textsubscript#1{%
  {\m@th\ensuremath{_{\mbox{\fontsize\sf@size\z@#1}}}}}
\makeatother

\usepackage[flushleft]{threeparttable}

\begin{document}

\title{Performance Modeling and Evaluation of Distributed Deep Learning Frameworks on GPUs}

\author{\IEEEauthorblockN{Shaohuai Shi, Qiang Wang, Xiaowen Chu}
\IEEEauthorblockA{Department of Computer Science, Hong Kong Baptist University
\\\{csshshi, qiangwang, chxw\}@comp.hkbu.edu.hk}
}

\maketitle

\begin{abstract}
Deep learning frameworks have been widely deployed on GPU servers for deep learning applications in both academia and industry. In training deep neural networks (DNNs), there are many standard processes or algorithms, such as convolution and stochastic gradient descent (SGD), but the running performance of different frameworks might be different even running the same deep model on the same GPU hardware. In this study, we evaluate the running performance of four state-of-the-art distributed deep learning frameworks (i.e., Caffe-MPI, CNTK, MXNet, and TensorFlow) over single-GPU, multi-GPU, and multi-node environments. We first build performance models of standard processes in training DNNs with SGD, and then we benchmark the running performance of these frameworks with three popular convolutional neural networks (i.e., AlexNet, GoogleNet and ResNet-50), after that, we analyze what factors that result in the performance gap among these four frameworks. Through both analytical and experimental analysis, we identify bottlenecks and overheads which could be further optimized. The main contribution is that the proposed performance models and the analysis provide further optimization directions in both algorithmic design and system configuration.
\end{abstract}


\IEEEpeerreviewmaketitle

\section{Introduction}
In recent years, deep learning (DL) techniques have achieved great success in many AI applications \cite{lecun2015deep}. With a large amount of data, deep neural networks (DNNs) can learn the feature representation very well. Very deep neural networks and large scale of data, however, result in a high requirement of computation resources. Fortunately, on one hand, GPUs play an important role in speeding up the training speed. On the other hand, it has been recently proven that DNNs with a very large size of mini-batch can converge well to a local minimal \cite{wang2017stochastic}\cite{goyal2017accurate}, which is significant to utilize many processors or clusters efficiently. A single accelerator has limited computational resources (e.g., computation units and memory) to process large-scale neural networks, so parallel training algorithms are proposed to solve this problem such as model parallelization \cite{lee2014model} and data parallelization \cite{zinkevich2010parallelized}\cite{chen2016revisiting}. Several popular distributed DL frameworks including Caffe-MPI\footnote{https://github.com/Caffe-MPI/Caffe-MPI.github.io}, CNTK\footnote{https://github.com/Microsoft/CNTK}, MXNet\footnote{https://github.com/apache/incubator-mxnet} and TensorFlow\footnote{https://github.com/tensorflow/tensorflow} have achieved not only high throughput in a single GPU with the help of cuDNN \cite{chetlur2014cudnn} which is a high performance DNN library provided by Nvidia, but they also have good scalability across multiple GPUs and multiple machines. These frameworks provide an easy way for users to develop DNNs and try to optimize related algorithms to achieve high throughput by using hardware platforms like multi-core CPU, many-core GPU, multiple GPUs and multiple machines. However, because of the different implementation methods by vendors, these tools show different performance even when training the same DNNs on the same hardware platform. Researchers have evaluated different tools on various hardware with diverse DNNs \cite{bahrampour2015comparative}\cite{shi2016benchmarking}\cite{shams2017evaluation}\cite{kim2017performance}, but the scalability, which is one of the most important factors in multi-GPU and multi-machine platforms, is not well studied. In this study, we extend the work in \cite{shi2016benchmarking} to evaluate the performance of four distributed DL frameworks (i.e., Caffe-MPI, CNTK, MXNet and TensorFlow) with convolutional neural networks (CNNs) over the GPU cluster. We use four machines connected by a 56Gbps InfiniBand network, each of which is equipped with four Nvidia Tesla P40 cards, to test the training speed of each framework in CNNs covering single-GPU, multi-GPU and multi-machine environments\footnote{Our source code and experimental data can be downloaded from \url{http://www.comp.hkbu.edu.hk/~chxw/dlbench.html}.}. We first build the performance models of SGD algorithm and then test the running performance of SGD optimization, and further focus on the performance of synchronous SGD (S-SGD) across multiple GPUs/machines to analyze the performance details. Our major findings are summarized as follows\footnote{The software tools are being upgraded frequently. The findings are based on our own experimental platforms, software configurations and only apply to the software versions specified in the paper.}: 
\begin{enumerate}
\item For relatively shallow CNNs (e.g., AlexNet), loading large amounts of training data could become a potential bottleneck with a large mini-batch size and fast GPUs. Efficient data pre-processing can be used to reduce the impact.
\item To better utilize cuDNN, autotune and input data layout (e.g., NCWH, NWHC) should be considered. Both CNTK and MXNet expose the autotune configuration of cuDNN, which could achieve better performance during the forward and backward propagation.
\item In S-SGD with multiple GPUs, CNTK does not hide the overhead of gradient communication, while MXNet and TensorFlow do by parallelizing the gradient aggregation of the current layer with the gradient computation of the previous layer. By hiding the overhead of gradient communication, the scaling performance could be better.
\item All the frameworks scale not so well across four high throughput dense GPU servers. The inter-node gradient communication via 56Gbps network interface is much slower than the intra-node via PCIe.
\end{enumerate}

The rest of the paper is organized as follows. Section \ref{backgroundandrelatedwork} introduces the related work. Section \ref{preliminaries} presents preliminaries of SGD and S-SGD implemented by different approaches. We derive some performance models for different implementations of S-SGD in Section \ref{modeling}. Our experimental methodology is introduced in Section \ref{methods}, followed by the experimental results and our analysis in Section \ref{results}. We conclude the paper and discuss our future work in Section \ref{conclusionandfuturework}.

\section{Background and Related Work} \label{backgroundandrelatedwork}
Stochastic gradient descent (SGD) methods are the most widely used optimizers in deep learning communities because of its good generalization and easy computation with the first order gradient \cite{rumelhart1988learning}\cite{zhang2016understanding}, and it can scale to multiple GPUs or machines for larger datasets and deeper neural networks. Distributed SGD methods have achieved good scaling performance \cite{chen2016revisiting}\cite{wang2017stochastic}\cite{goyal2017accurate}, and the existing popular DL frameworks have the built-in components to support scaling by using some configurations or APIs, among which Caffe-MPI, CNTK, MXNet and TensorFlow are examples of the most active and popular ones. However, these frameworks implement the working flow of SGD in different ways, which results in some performance gap even though they all make use of high performance library cuDNN \cite{chetlur2014cudnn} to accelerate the training on GPUs. In addition, the implementation of S-SGD may vary so much for different purposes. 

Parameter server (PS) based methods \cite{li2014scaling}\cite{cui2016geeps} for distributed machine learning algorithms have been widely used in many distributed SGD algorithms like asynchronous SGD \cite{zhang2013asynchronous} and S-SGD. Several performance models for PS methods have been proposed by S. Zou et al. \cite{zou2017distributed}, and they develop the procedure to help users better choose the mini-batch size and the number of parameter servers. 

A. Awan et al. \cite{awan2017s}\cite{awan2017optimized} propose the high performance CUDA-Aware MPI to alleviate the overhead of data communication so that they can scale the distributed learning better on GPU clusters. In the recent research, P. Goyal et al. \cite{goyal2017accurate} use a dense GPU cluster with 256 GPUs to achieve about 90\% efficiency. Except the PS-based method used in \cite{goyal2017accurate}, the optimized all-reduce implementation and pipelining all-reduce operations with gradient computation make training nearly perfect linear scale up possible in ResNet-50 \cite{he2015deep}. Most of these researches focus on the optimization of PS-based methods which have very high requirement on the network bandwidth between the parameter server and workers, while the decentralized methods are less studied since they are initially considered to highly rely on the PCIe topology between GPU and CPU. X. Lian et al. \cite{lian2017can} come up with a decentralized S-SGD algorithm which has a theoretical guarantee of convergence to overcome the communication bottleneck across the dense GPU cluster. Even though this work only conducts experiments on the small size of the dataset, it lets us re-consider the importance of decentralized S-SGD algorithms on GPU clusters. And the hybrid method of PS-based and decentralized is also proposed to speed up training \cite{zhang2017poseidon}. It is noted that both PS-based methods and the decentralized S-SGD have been integrated into most distributed DL frameworks. 

Bahrampour et al. \cite{bahrampour2015comparative} and Shi et al. \cite{shi2016benchmarking} have evaluated the performance of some state-of-the-art DL frameworks on the single-GPU environment. But they did not break down the timing of the training process, which lacks details to understand performance problems. In the distributed environment, Shams et al. \cite{shams2017evaluation} have studied the performance of Nvidia's NVLink and Intel\rq s Knights Landing on different CPU and GPU technologies. However, the evaluated TensorFlow is at version v0.12, while Google has upgraded TensorFlow to v1.0+ for performance improvement, and the other two popular commercial frameworks (CNTK and MXNet) are not compared in \cite{shams2017evaluation}. In addition, the performance model in the distributed GPU cluster is also not studied. In this paper, we first build performance models of SGD in both the single node and the distributed cluster, and then compare the performance of Caffe-MPI, CNTK, MXNet and TensorFlow via single-GPU, multi-GPU and multi-node environments through analysis and experimental results, and then identify the performance gap among these four frameworks.

\section{Preliminaries} \label{preliminaries}
In this section, we first introduce the workflow of SGD and S-SGD, and then we illustrate the current implementations of S-SGD. Some frequently used notations in our analysis are summarized in Table \ref{table:notation}. We assume that each node in the cluster has the same hardware configuration.

\begin{table}[!ht]
	\centering
	\caption{Summary of notations}
	\label{table:notation}
	\begin{tabular}{|l|l|}
		\hline
		Name &  Description \\\cline{1-2}
		\hline
		\hline
		$N_g$ & \# of total GPUs \\\cline{1-2}
		$n_g$ & \# of GPUs on each node \\\cline{1-2}
		$M$ & \# of training samples per GPU in a mini-batch \\\cline{1-2}
		$t_{iter}$ & Time of an iteration\\\cline{1-2}
		$t_{io}$ & Time of I/O in each iteration\\\cline{1-2}
		$t_{h2d}$ & Data transfer time from CPU to GPU in each iteration\\\cline{1-2}
		$t_{f}$ & Time of the forward phase in each iteration\\\cline{1-2}
		$t_{b}$ & Time of the backward phase in each iteration\\\cline{1-2}
		$t_{b}^{(i)}$ & Time of the backward phase of layer $i$ in each iteration\\\cline{1-2}
		$t_{u}$ & Time of the model update in each iteration\\\cline{1-2}
		$t_{comm}$ & Time of the gradients aggregation in each iteration\\\cline{1-2}
		$t_{comm}^{(i)}$ & Gradients aggregation time of layer $i$ in each iteration\\\cline{1-2}
	\end{tabular}
	\vspace{-10pt}
\end{table}

\subsection{Mini-batch SGD} \label{sgd}
To train a model with mini-batch SGD, one should update the model iteratively with feeding data. It generally contains five steps in an iteration. 1) Read a mini-batch of data from the disk to the memory. 2) Transfer the data from the CPU memory to the GPU memory. 3) Launch GPU kernels to do feed forward operations layer by layer. 4) Do backward propagation by calculating first order gradients w.r.t weights and inputs with the chain rule. 5) Update the model by gradients. So the total time of one iteration can be calculated as
\begin{equation}
t_{iter}=t_{io}+t_{h2d} + t_{f} + t_{b} + t_{u}.
\end{equation}

\subsection{S-SGD} \label{ssgd}
In general, S-SGD makes each worker perform feed forward and backward propagation with different samples and a duplicate copy of the model. Before updating the model, the gradients are aggregated \cite{zinkevich2010parallelized}. There are five steps to implement the naive S-SGD algorithm with a distributed cluster. 1) Each machine reads and/or preprocesses $M\times n_g$ samples, and it totally has $N\times M\times n_g$ samples for $N$ nodes. 2) In each machine, $M\times n_{g}$ samples are evenly distributed to $n_{g}$ different GPUs through PCIe. 3) Each GPU launches kernels to do feed forward and backward propagation operations in parallel. 4) The gradients are averaged among all the GPUs. 5) Each GPU updates its own parameters. In step 4), the aggregation operation should wait for all the GPUs sending the gradients of that iteration, which indicates the meaning of synchronous SGD. 



The PS method \cite{li2014scaling} is one of the state-of-the-art methods to implement S-SGD. It is a centralized topology. There is a parameter server (PS) to store the whole model in a single node, and it can be extended to two or more PSes if needed. PS aggregates parameters at each iteration and updates the model and then pushes the updated model to each worker. As a centralized node, it may easily suffer from the high pressure if the number of parameters is huge.

The decentralized method is another algorithm to implement the gradients aggregation by using the reduction tree (RT) \cite{awan2016efficient}\cite{awan2017s}. The gradients are exchanged via MPI-like collectives (e.g., all-reduce). So there come out some efficient collective libraries like Gloo\footnote{https://github.com/facebookincubator/gloo} and NCCL2\footnote{https://developer.nvidia.com/nccl} supporting communication between distributed GPUs.

\section{Performance Modeling} \label{modeling}
In this section, we build the performance models of training DNNs with SGD (or S-SGD) in Caffe-MPI, CNTK, MXNet and TensorFlow. From the workflow described in Sections \ref{sgd} and \ref{ssgd}, it is straightforward to represent the iteration time $t_{iter}$ with:
\begin{equation}\label{seqiter}
t_{iter} = t_{io} + t_{h2d} + t_{f} + t_{b} + t_{comm} + t_{u}.
\end{equation}
Let $t_{gpu}=t_{h2d} + t_{f} + t_{b} + t_{u}$, then we have
\begin{equation}\label{seqitersimple}
t_{iter} = t_{io} + t_{gpu} + t_{comm}.
\end{equation}
In the single-GPU environment, $t_{comm}=0$. $t_{io}$ and $t_{comm}$ in Equation \ref{seqitersimple} can be hidden to some extent by pipeline techniques.

\subsection{I/O hidden}
To achieve higher efficiency of training, step 1) is often processed with multiple threads so that the I/O time of a new iteration can be overlapped with the computing time of the previous iteration. Data can be accessed from the CPU memory directly without waiting for the data from the disk if it has been ready during the computation of the previous iteration. So we can calculate the average iteration time of pipelined SGD as
\begin{equation}
\label{equ:pipeioiter}
\bar{t}_{iter} = max\{t_{gpu} + t_{comm}, t_{io}\}.
\end{equation}


\subsection{Communication hidden}
The main property of the mini-batch SGD training of CNNs is that the gradient computation has no dependency with the updating of their next layers, so the gradient computation in layer $l_{i-1}$ can be parallelized with the gradient aggregation in layer $l_i$ \cite{zhang2017poseidon}\cite{awan2017s}. Let $\tau_{comm\_s}$ and $\tau_{comm\_e}$ denote the start and the end timestamps of communication during one iteration respectively. The iteration time can be represented by 
\begin{equation}
t_{iter} = t_{io} + t_{h2d} + t_{f} + t_{b}^{(L)} + \tau_{comm\_e} - \tau_{comm\_s} + t_{u}\text{,}
\end{equation}
where $t_{b}^{(L)}$ is the gradient computation time of the last learnable layer. We discuss two cases:

{\textbf{Case 1.}} The gradient communication is totally hidden by the backward propagation. I.e., $t_{comm}^{(i)} \leq t_{b}^{(i-1)}, 2\leq i \leq L$. We can update the representation of $t_{iter}$ by:
\begin{equation}\label{equ:optiter}
t_{iter} = t_{io} + t_{h2d} + t_{f} + t_{b} + t_{comm}^{(1)} + t_{u}^{(1)}\text{.}
\end{equation}

{\textbf{Case 2.}} There exist some layers whose communication time are longer than the time of backward computation of the previous layers. We formulate this case with: $t_{comm}^{(i)} \leq t_{b}^{(i-1)}$ for $i=2, ..., C-1$, and $t_{comm}^{(i)} > t_{b}^{(i-1)}$ for $i=C, C+1, ..., L$. Thus, $t_{iter}$ is estimated by
\begin{equation}\label{equ:commbottleneckiter}
t_{iter} = t_{io} + t_{h2d} + t_{f} + \sum_{i=C}^{L}t_{comm}^{(i)} + \sum_{i=1}^{C-1}t_{b}^{(i)}+ t_{comm}^{(1)} + t_{u}^{(1)}\text{,}
\end{equation}
where $L$ is the number of learnable layers of DNN. It is noted the larger $C$, the more communication can be hidden. 

Let $t_{iter\_N_g}$ and $t_{io\_n_g}$ denote the iteration time and the I/O time of a mini-batch with $N_g$ GPUs across $N$ machines (each machine has $n_g$ GPUs) respectively. The speedup can be formulated by
\begin{equation}\label{equ:speedup}
\begin{split}
S &= \cfrac{MN_{g}/t_{iter\_N_{g}}}{M/t_{iter\_1}}=N_g\frac{t_{io\_1} + t_{gpu} }{t_{io\_n_g} + t_{gpu} + t_{comm}}.
\end{split}
\end{equation}
So to achieve good scalability of the system, one should reduce the overheads of I/O and data communication. 

For CNTK, the speedup of S-SGD is estimated by Equation \ref{equ:speedup}, while for the tools (Caffe-MPI, CNTK and TensorFlow) that exploit the pipelining between backward and communication, the speedup can be estimated by
\begin{equation}\label{equ:speedupmxnet1}
S= \\
\begin{cases}
\frac{N_g(t_{io\_1} + t_{gpu})}{t_{io\_n_g} + t_{h2d} + t_{f} + t_{b} + t_{comm}^{(1)} + t_{u}^{(1)}} & \text{\textbf{Case 1}}\\
\frac{N_g(t_{io\_1} + t_{gpu})}{t_{io\_n_g} + t_{h2d} + t_{f} + \sum_{i=C}^{L}t_{comm}^{(i)} + \sum_{i=1}^{C-1}t_{b}^{(i)} + t_{comm}^{(1)} + t_{u}^{(1)}} & \text{\textbf{Case 2}}\\
\end{cases}.
\end{equation}

\section{Experimental Methodology} \label{methods}
\begin{figure}[!ht]
	\centering
	\includegraphics[width=0.8\linewidth]{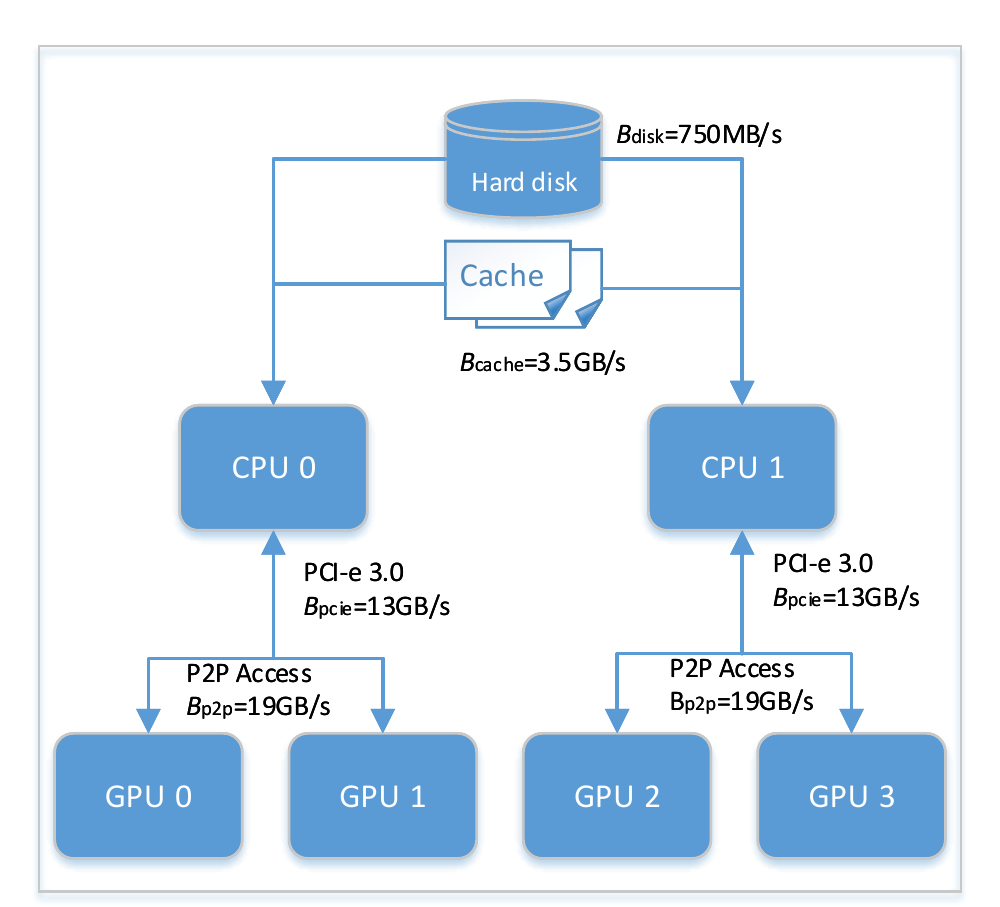}
	\caption{The topology of a single node. \textit{Cache} refers to the system cache that contains parked file data. The values of $B_{disk}$ and $B_{cache}$ are benchmarked by \textit{dd} command. The bandwidth of PCIe and P2P access are measured via Nvidia CUDA SDK samples.}
	\label{fig:node}
\end{figure}

We first specify the hardware environment conducted in the experiments. We use a 4-node GPU cluster, in which each node has four Nvidia Tesla P40 cards, and the network connection between nodes is a 56 Gbps InfiniBand combined with a 1 Gbps Ethernet. Table \ref{table:multigpuetup} shows the hardware setting. The intra-node topology with a bandwidth of data transfer between different components is displayed in Fig. \ref{fig:node}. Each Tesla P40 GPU runs at the base core frequency of 1.3 GHz and the auto boost function is disabled to ensure reproducibility of our experimental results.

\begin{table}[!ht]
	\centering
	\caption{The experimental hardware setting for data parallelization.}
	\label{table:multigpuetup}
	\begin{tabular}{|l|l|}
		\hline
		Hardware& 	Model     \\\hline
		\hline
		GPU  		& 	Nvidia Tesla P40      \\\hline
		CPU	 		&	Intel Xeon E5-2650v4 Dual	\\\hline
		Network		&	56 Gbps InfiniBand + 1 Gbps Ethernet \\\hline
		Memory		&	128 GB DDR4 	\\\hline
		Hard disk	&	SSD 6T (x2 with RAID 0) in Node 0, \\
					&and others are HDD 1T (x4 with RAID 5).\\
					& Each node has one copy of the dataset \\\hline
	\end{tabular}
\end{table}

Versions of the tested frameworks installed in each node are shown in Table \ref{table:software_new}. The operating system of the server is CentOS 7.2, and the software is installed with CUDA-8.0 and cuDNNv6.

\begin{table}[!ht]
	\centering
	\caption{The softwares used for experiments.}
	\label{table:software_new}
	\begin{tabular}{|l|l|l|}
		\hline
		Software  &  Marjor Version & GitHub Commit ID  \\
		\hline\hline
		Caffe-MPI      &  2.0	  	& -   \\\hline
		CNTK      &  2.1	  	& 4a8db9c   \\\hline
		MXNet 	  &  0.10.0 	& 34b2798   \\\hline
		TensorFlow & 1.2.1		& -         \\\hline
	\end{tabular}
	\vspace{-10pt}
\end{table}

One popular and effective way to evaluate the running performance is to measure the time duration of an iteration that processes a mini-batch of input data or the number of samples can be processed in one second. Therefore, we benchmark the CNNs by using a proper mini-batch size (try to fully utilize the GPU resource) for each network on these tools.

We choose three popular CNNs (i.e., AlexNet \cite{krizhevsky2012imagenet}, GoogleNet \cite{szegedy2015going} and ResNet-50 \cite{he2015deep}) running on the ILSVRC-2012 ImageNet dataset \cite{deng2009imagenet}. These three deep models have their own characteristics to test the performance of frameworks. They have different configurations and the details are shown in Table \ref{table:networksetup}. Each machine in the cluster has one copy of the dataset. The data formats for different frameworks are not the same, and we list the data formats below for the tested frameworks. \textbf{Caffe-MPI}: LMDB is used by Caffe-MPI. The original JPEG images are converted to LMDB records, and the script can be found in the GitHub repository of Caffe\footnote{\url{https://github.com/BVLC/caffe/blob/master/examples/imagenet/create_imagenet.sh}}.
\textbf{CNTK}: There is no pre-converted data format for CNTK. It needs to read the original JPEG images during training with a provided file list. \textbf{MXNet}: A binary file that contains all the images is used. The converting script refers to the official document of MXNet\footnote{\url{https://github.com/apache/incubator-mxnet/tree/master/example/image-classification#prepare-datasets}}. \textbf{TensorFlow}: It also uses a pre-converted file format called \textit{TFRecord} in TensorFlow. The converting script refers to the script from the GitHub repository\footnote{\url{https://github.com/tensorflow/models/blob/master/research/inception/inception/data/build_imagenet_data.py}}.

\begin{table}[htbp]
\begin{threeparttable}
\centering
\caption{The experimental setup of neural networks.}
\label{table:networksetup}
\begin{tabular}{|l|l|l|l|l|}
	\hline
Network & \# of Layers & \# of FCs & Parameters  & Batch size \\\hline\hline
AlexNet   & 8     & 3  & \textasciitilde 60 millions   & 1024  \\\hline
GoogleNet & 22    & 1  & \textasciitilde 53 millions & 128 \\\hline
ResNet-50 & 50 	  & 1  & \textasciitilde 24 millions & 32 \\\hline
\end{tabular}
	\begin{tablenotes}
      \item[]Note: The architecture of AlexNet is the same with \cite{krizhevsky2012imagenet} except that the local response normalization (LRN) operation is excluded because it is not supported by CNTK by default. We choose the proper batch size for each device, which can be run properly for all frameworks, and it tries to fully utilize the GPU resource.
    \end{tablenotes}
  \end{threeparttable}
\end{table}

In order to avoid the heavy impact of I/O overheads from hard disks, we run two epochs and the first epoch is excluded to calculate the average time of one iteration. Since the total number of images is up to 1.2 million, it is very time-consuming to run all the samples in one epoch. So we limit the epoch size to make each experiment run about 50-100 batches in one epoch. The time of each iteration is recorded and all iterations in the second epoch are averaged to calculate the mean and standard deviation to measure the running performance.

Beside the running speed measured in this paper, we also break down the timing for each phase by using \textit{nvprof}, which is a tool to profile the GPU activities, to help us identify performance problems.

\section{Experimental Results and Analysis} \label{results} 
In this section, we demonstrate the running performance followed with analysis based on the modeling of CNTK, MXNet and TensorFlow in training AlexNet, GoogleNet and ResNet-50 on a single P40 card, multiple P40 cards, and across the 4-node GPU cluster.

\subsection{Single GPU}
We first present the performance results on a single GPU. The average time of one iteration during training is used to metric the performance of frameworks. So we compare the time cost in each step of SGD. We break down the timing of each phase in Table \ref{table:sgbreakdown}. Results in each phase will be analyzed in the following.


\begin{table}[!ht]
	\centering
	\caption{The time breakdown of different phases of SGD in second ($mean\pm std$).}
	\label{table:sgbreakdown}
\bgroup
\setlength\tabcolsep{2pt}
	\begin{tabular}{|l|l|l|l|l|}
		\hline
\textbf{AlexNet}    &Caffe-MPI         &CNTK              &MXNet             &TensorFlow        \\\hline
$t_{io}$   &.0002$\pm$6.9e-05 &.2233$\pm$5.1e-02 &.0001$\pm$1.8e-05 &.0008$\pm$9.3e-04 \\\hline 
$t_{h2d}$  &.0526$\pm$3.4e-04 &.0528$\pm$4.1e-04 &.1109$\pm$1.6e-02 &.1140$\pm$1.5e-02 \\\hline 
$t_{f}$    &.1718$\pm$8.4e-03 &.1684$\pm$1.9e-03 &.2147$\pm$3.5e-04 &.1804$\pm$1.2e-02 \\\hline 
$t_{b}$    &.3560$\pm$5.6e-03 &.2919$\pm$9.3e-04 &.4086$\pm$1.5e-03 &.3417$\pm$3.1e-02 \\\hline 
$t_{u}$    &.0062$\pm$1.0e-05 &.0086$\pm$2.3e-06 &.0041$\pm$4.5e-06 &.0031$\pm$1.7e-06 \\\hline 
$t_{iter}$ &\textbf{.5772$\pm$6.0e-02} &.7433$\pm$1.0e-02 &.7235$\pm$1.0e-01 &.6593$\pm$1.6e-02 \\\hline\hline

\textbf{GoogleNet}  &Caffe-MPI         &CNTK              &MXNet             &TensorFlow        \\\hline 
$t_{io}$   &.0002$\pm$3.8e-04 &.0001$\pm$8.2e-06 &.0001$\pm$2.1e-05 &.0010$\pm$4.6e-04 \\\hline 
$t_{h2d}$  &.0021$\pm$2.6e-03 &.0033$\pm$3.3e-03 &.0143$\pm$1.8e-03 &.0161$\pm$1.5e-03 \\\hline 
$t_{f}$    &.0920$\pm$1.8e-03 &.0814$\pm$7.5e-04 &.0892$\pm$1.2e-04 &.1192$\pm$2.3e-02 \\\hline 
$t_{b}$    &.2073$\pm$4.0e-04 &.1780$\pm$1.4e-03 &.1819$\pm$2.1e-04 &.1856$\pm$2.4e-04 \\\hline 
$t_{u}$    &.0015$\pm$1.2e-05 &.0095$\pm$1.7e-03 &.0095$\pm$1.5e-05 &.0005$\pm$5.6e-06 \\\hline 
$t_{iter}$ &.3050$\pm$1.3e-02 &\textbf{.2767$\pm$2.0e-02} &.2943$\pm$8.4e-03 &.3100$\pm$1.5e-03 \\\hline\hline

\textbf{ResNet}     &Caffe-MPI         &CNTK        &MXNet                   &TensorFlow        \\\hline
$t_{io}$   &.0002$\pm$5.4e-05 &.0001$\pm$1.1e-05 &.0001$\pm$9.9e-06 &.0003$\pm$1.2e-04 \\\hline 
$t_{h2d}$  &.0009$\pm$7.2e-04 &.0016$\pm$2.7e-05 &.0039$\pm$5.7e-04 &.0045$\pm$1.0e-03 \\\hline 
$t_{f}$    &.0807$\pm$4.3e-05 &.0724$\pm$1.0e-03 &.0732$\pm$5.4e-05 &.0767$\pm$1.2e-02 \\\hline 
$t_{b}$    &.1291$\pm$5.4e-03 &.1482$\pm$9.4e-04 &.1307$\pm$1.4e-04 &.1355$\pm$1.4e-04 \\\hline 
$t_{u}$    &.0033$\pm$4.5e-05 &.0073$\pm$4.7e-03 &.0164$\pm$2.5e-05 &.0015$\pm$1.8e-06 \\\hline 
$t_{iter}$ &\textbf{.2078$\pm$5.8e-03} &.2261$\pm$1.4e-02 &.2242$\pm$1.2e-02 &.2148$\pm$3.4e-02 \\\hline
	\end{tabular}
	\egroup
\end{table}

\textbf{I/O}. Among the evaluated tools, they all support data prefetch, which means during training, there are extra threads reading data to the CPU memory for preparing to feed into GPU. However, some implementation details are not the same. Regarding Caffe-MPI, there exist GPU buffers to prefetch the data, which means that each iteration, except the first one, can load data from the GPU memory without waiting for I/O and PCIe transfer. For CNTK, there could be a limited buffer for data caching, which may result in the dropdown of performance if the size of data in a mini-batch is too large. On the contrary, MXNet and TensorFlow are more flexible and have little opportunity to fall into I/O problem. In Table \ref{table:sgbreakdown}, CNTK has a big overhead in reading data when running AlexNet with a mini-batch size of 1024. The reason is that it needs $S_{d} = 1024\times 224 \times 224\times3\times4=588$MB to store data of one batch, while it is not fast enough to store data of next batch in CNTK. CNTK needs to read and decode the original JPEG files to prepare the input data while other frameworks just need to read from pre-converted files. From Fig. \ref{fig:node}, the bandwidth of system cache is $B_{cache}=3.5$GB/s, so the overhead of reading data is $t_{io}=\frac{S_{d}}{B_{cache}}=\frac{588}{(3.5\times1024)}s=0.164$s, which is the optimal time, adding the time of decoding JPEG files, the actual value of $t_{io}$ is $0.223$s in CNTK. By contrast, MXNet and TensorFlow only need negligible time in reading data.

\textbf{Memory copy: from host to device (h2d)}. After reading data from disk to memory, data should be transferred to GPU for training. In our tested environment, CPU and GPU are connected by PCIe with a bandwidth of 13 GB/s. It is noticed that $t_{h2d}$ in both Caffe-MPI and CNTK are about half smaller than MXNet and TensorFlow even the size of data is same because of the difference in memory allocation. There are non-pageable and pageable memories, and their performances of memory copy from CPU to GPU are different \cite{nvidia2011nvidia}. The bandwidth of non-pageable and pageable memory copy in the tested hardware is 11.4 GB/s and 8.7 GB/s respectively. Since Caffe-MPI and CNTK allocate the non-pageable memory for input data, while MXNet and TensorFlow allocate pageable memories, so CNTK achieves better memory copy performance than that of MXNet and TensorFlow.

\textbf{Forward, backward and update}. The high performance library cuDNN \cite{chetlur2014cudnn} on GPUs, provided by Nvidia, has been widely used in DL frameworks. During the training of DNNs, most of the time-consuming layers (e.g., convolutional) are invoked by cuDNN. However, parameters in APIs of cuDNN may result in different performances, which is the main reason why $t_f$ and $t_b$ are different in Table \ref{table:sgbreakdown}. For example, there are many types of implementations of convolution like GEMM, FFT and Winograd. Users can specify which algorithm to use or autotune to choose the best one. When invoking the APIs of cuDNN, another performance-related factor is the data layout (e.g., NCWH, NWHC). In both forward and backward phases, CNTK achieves the best performance in all networks. Actually, Caffe-MPI, CNTK and MXNet could autotune to find the best convolution algorithms for convolutional layers, but TensorFlow prefers to use Winograd algorithm which could be suboptimal in some cases. Regarding AlexNet, CNTK invokes the FFT algorithm for the second convolutional layer, while MXNet uses the GEMM-based convolution so that there is 0.04s larger in the forward phase and up to 0.1s higher in the backward phase. The FFT-based convolution is faster than the GEMM-based in general \cite{mathieu2013fast}. The suboptimal invoking of cuDNN APIs makes TensorFlow slightly worse than CNTK in both forward and backward phases. The update operation is simple since it only updates the parameters with computed gradients, and the time complexity is O(1), whose time cost is relatively short compared to the forward and the backward propagations. But CNTK and MXNet perform not that good in this phase.

In summary, CNTK has faster data copy, forward and backward propagation operations, which results in better performance in GoogleNet compared to MXNet and TensorFlow. Caffe-MPI outperforms CNTK in AlexNet since Caffe-MPI can hide the overhead of I/O. However, the test of GoogleNet has two advantages for CNTK. First, there are many large size of filters (e.g., $5\times5$) in convolutional layers, which could reflect the importance of convolution algorithm selection. Second, the mini-batch size used is only 128, such that it has a very small overhead in data loading. MXNet and TensorFlow have better data prefetch mechanism than CNTK. It is obvious in the case of AlexNet, though TensorFlow has a suboptimal performance in data copy, forward and backward, it hides the overhead of data reading. As a result, TensorFlow achieves 10\% faster than CNTK in AlexNet. Regarding ResNet-50, convolutional layers are with smaller kernels (e.g., $3\times3$) which requires less computation and the Winograd algorithm could be the better implementation \cite{lavin2016fast}.

\subsection{Multiple GPUs}
\begin{figure}[!ht]
	\centering
	\includegraphics[width=\linewidth]{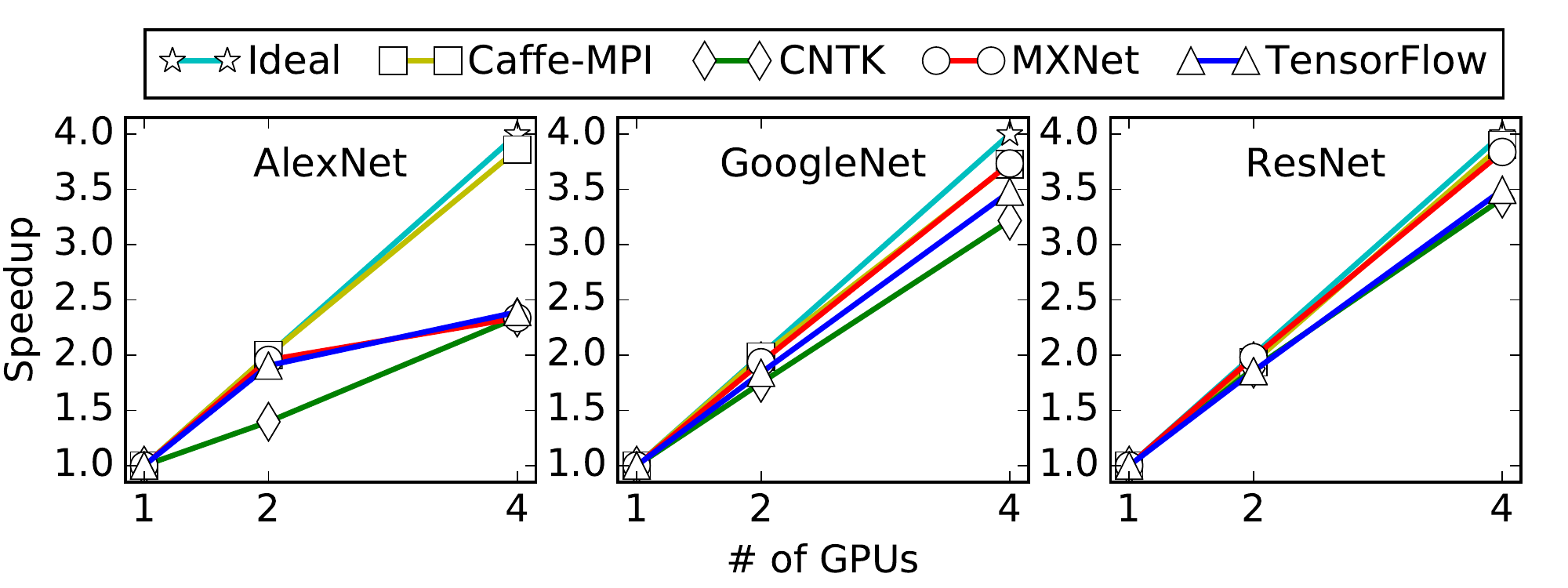}
	\vspace{-20pt}
	\caption{Scaling performance in a single node.}
	\label{fig:multiplegpus}
\end{figure}

When scaling to multiple GPUs, it is important to shorten the overhead of data aggregation, which heavily relies on the bandwidth of data communication between GPUs. Please be noted that in multi-GPU/node testing we use weak scaling, which means the valid mini-batch size is scaling with the number of GPUs, and each GPU keeps the same number of samples like the work in \cite{wang2017stochastic}\cite{goyal2017accurate}. To reflect the scalability of deep learning frameworks, we use the metric of samples per second to compare the performance. Ideally, the throughput should be doubled with the number of GPUs doubled. The scaling performance of S-SGD running on a machine with four GPUs is shown in Fig. \ref{fig:multiplegpus}. Let $t_{comm}$ denote the overhead of gradient communication when aggregating the gradients, and the numbers are shown in Table \ref{table:mgtcomm}.

\begin{table}[!ht]
	\centering
	\caption{The overhead of data communication when aggregating the gradients across intra-node multiple GPUs.}
	\label{table:mgtcomm}
	\setlength\tabcolsep{4pt}
	\begin{tabular}{|l|l|l|l|l|}
		\hline
	Network		&	Tool				& \multicolumn{2}{c|}{$t_{comm}$} & Hidden \\\cline{3-4}
						&			& 2 GPUs & 4 GPUs & \\\hline\hline
&Caffe-MPI       & .0457$\pm$3.1e-03 & .0861$\pm$1.2e-02  & Yes  \\\cline{2-5}
&CNTK            & .0359$\pm$2.4e-02 & .0420$\pm$3.0e-03 & No  \\\cline{2-5}
AlexNet &MXNet   & .0222$\pm$4.5e-03 & .0505$\pm$7.7e-03 & Yes \\\cline{2-5}
&TensorFlow      & .0406$\pm$8.9e-03 & .0984$\pm$1.9e-02 & Yes  \\\hline
&Caffe-MPI       & .0135$\pm$1.4e-03 & .0229$\pm$1.2e-03  & Yes  \\\cline{2-5}
&CNTK            & .0343$\pm$1.2e-02 & .0592$\pm$6.2e-03 & No  \\\cline{2-5}
GoogleNet &MXNex & .0102$\pm$8.9e-04 & .0318$\pm$1.1e-02 & Yes \\\cline{2-5}
&TensorFlow      & .0053$\pm$3.8e-04 & .0168$\pm$1.6e-03 & Yes  \\\hline
&Caffe-MPI       & .0336$\pm$2.4e-03 & .0646$\pm$2.9e-03   & Yes  \\\cline{2-5}
&CNTK            & .0173$\pm$1.2e-02 & .0295$\pm$2.1e-02 & No  \\\cline{2-5} 
ResNet-50 &MXNex & .0491$\pm$4.1e-02 & .1626$\pm$1.1e-01 & Yes \\\cline{2-5}
&TensorFlow      & .0072$\pm$2.1e-03 & .0210$\pm$5.0e-03 & Yes  \\\hline
	\end{tabular}
\end{table}

From Fig. \ref{fig:multiplegpus} (a), we can see that Caffe-MPI, MXNet and TensorFlow have achieved almost linear scaling from one to two GPUs, while CNTK has only a slight speedup with multiple GPUs. Caffe-MPI, MXNet and TensorFlow parallelize the gradient aggregation with backward propagation. In other words, the previous layer ($l_{i-1}$) of backward propagation can happen without any delay after gradients of current layer ($l_i$) computed, and at this time, gradient computation of $l_{i-1}$ is parallelized with gradient aggregation of $l_{i}$. In this way, much of the synchronization overhead of early gradients can be hidden by later computation layers. From Table \ref{table:mgtcomm}, it is noted that Caffe-MPI, MXNet and TensorFlow can hide $t_{comm}$ while CNTK does not. CNTK processes gradient computation and aggregation in a sequential way. Fortunately, the overhead of gradient aggregation can be highly reduced by high performance all-reduce library NCCL which is used by CNTK. 

Regarding AlexNet, the low scaling efficiency of CNTK is caused by the data reading from the disk to the memory. Since the data buffer is not fast enough to prefetch the next-batch data, the GPU computation needs to wait for the data loading from the disk to the memory in every iteration. Suppose that the data of one epoch has been loaded into the system cache, and the data size is up to $S_d=588N_g$ MB, we have $t_{io}=\frac{S_d}{B_{cache}}=\frac{588\times N_g}{3.5\times 1024}=0.164N_g$. In the tested cases of CNTK on AlexNet, $t_{io}$ is up to 0.45s and 0.72s with 2 and 4 GPUs respectively so that CNTK has a poor scaling performance with 2 GPUs and 4 GPUs. From Table \ref{table:sgbreakdown}, we have $t_{gpu}=0.0527+0.1684+0.2918+0.0086=0.5215$. According to Equation \ref{equ:speedup}, $S=\frac{N_g(0.223+0.5215)}{0.5215+t_{io\_N_g}+t_{comm}}$. For 2 GPUs, $S=\frac{2\times 0.7445}{0.5215+0.45+0.0359}=1.478$, and for 4 GPUs, we have $S=\frac{4\times 0.7445}{0.5215+0.72+0.042}=2.32$. The estimated speedup can match the evaluated results in Fig. \ref{fig:multiplegpus}. There is a similar scenario on 4-GPU training of AlexNet (0.55s for data reading) using MXNet and TensorFlow. In S-SGD with multiple GPUs in a single machine, every data reading thread fetches 4096 samples and distributes them to 4 GPUs. Since $t_{io}$ is longer than $t_{gpu}$, the I/O time cannot be hidden totally, which results in poor scaling across 4 GPUs.

For GoogleNet and ResNet-50, CNTK achieves worse scaling performance than Caffe-MPI, MXNet and TensorFlow since CNTK does not parallelize the gradient computation and aggregation. Since the overhead of I/O can be hidden this time, according to Equation \ref{equ:speedup}, gradients aggregation becomes the main factor that influences the scaling performance in CNTK. MXNet achieves better scaling efficiency than TensorFlow. In MXNet, there is a parameter server (on GPU) to keep a copy of parameters. When gradients of layer $l_i$ have been calculated, the gradients from multiple GPUs are transferred to PS, and then PS aggregates gradients and updates the model parameters directly, and then it copies parameters back to all GPUs. In this way, MXNet can hide both overheads of gradient synchronization and model updating. By contrast, TensorFlow implements S-SGD in a different way. It has no PS, and it uses peer-to-peer memory access if the hardware topology supports it. Beside the decentralized method, another main difference is that each GPU needs to average the gradients from other GPUs and updates the model after the backward propagation finished. Therefore, that the model updating is not overlapped with backward propagation leads to suboptimal scaling performance of TensorFlow. 

In conclusion, two things are important to reduce the impact of gradient aggregation in S-SGD. On one hand, high speed of data communication between GPUs is important to ease the overhead of gradients synchronization. On the other hand, the parallelism between communication and computation is necessary to hide the overhead of communication.

\subsection{Multiple machines}
\begin{figure}[!h]
	\centering
	\includegraphics[width=\linewidth]{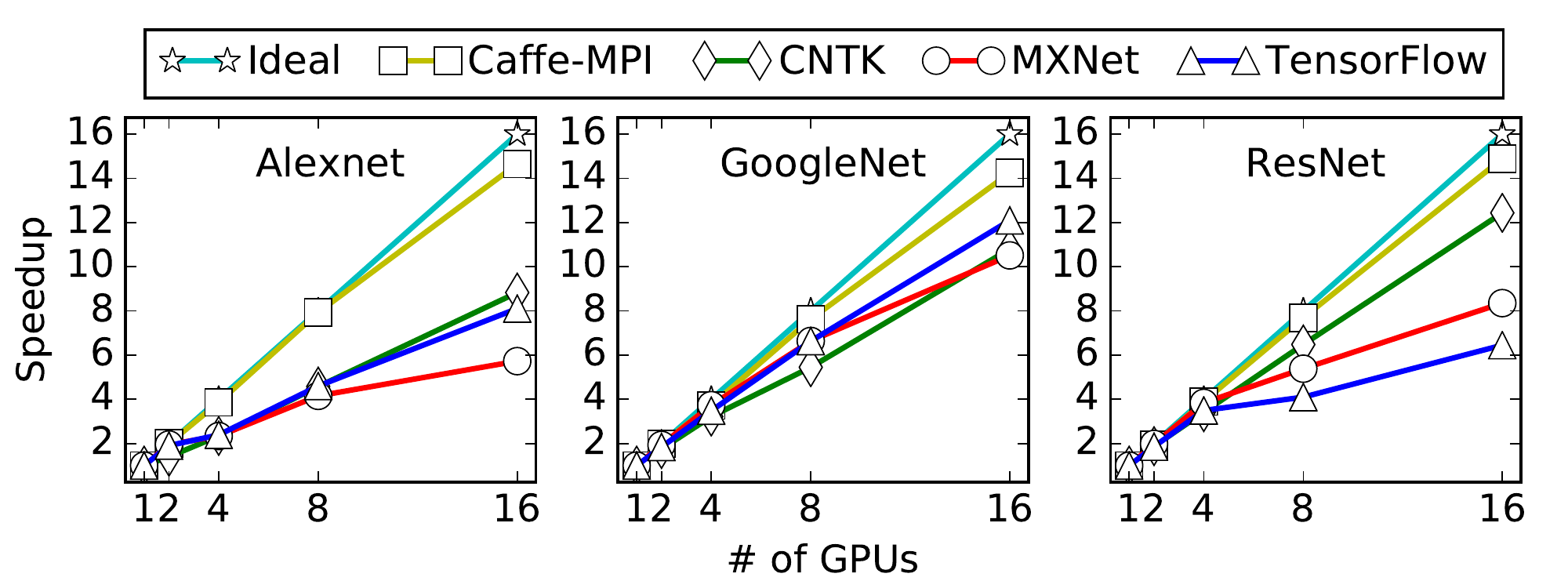}
	\vspace{-20pt}
	\caption{Scaling performance with multiple machines (each machine has 4 GPUs).}
	\label{fig:multilenodes}
\end{figure}
\begin{table}[!h]
	\vspace{-10pt}
	\centering
	\caption{The overhead of data communication when aggregating the gradients across multiple machines.}
	\label{table:mntcomm}
	\setlength\tabcolsep{4pt}
	\begin{threeparttable}
	\begin{tabular}{|l|l|l|l|l|}
		\hline
		Network		&	Tool	& \multicolumn{2}{c|}{$t_{comm}$} & Hidden  \\\cline{3-4}
		&			& 2 nodes & 4 nodes & \\\hline\hline
&Caffe-MPI       & .1344$\pm$6.6e-03 & .1650$\pm$1.9e-02 & Yes  \\\cline{2-5}
&CNTK            & .0906$\pm$9.7e-03 & .2364$\pm$7.5e-02 & No  \\\cline{2-5}
AlexNet   &MXNet & .5004$\pm$2.5e-02 & .7513$\pm$5.6e-01 & No \\\cline{2-5}
&TensorFlow      &    -              &  -                & No  \\\hline
&Caffe-MPI       & .1161$\pm$4.7e-02 & .0867$\pm$3.0e-02 & Yes  \\\cline{2-5}
&CNTK            & .1032$\pm$3.6e-02 & .1105$\pm$1.5e-02 & No  \\\cline{2-5}
GoogleNet &MXNet & .2973$\pm$1.9e-01 & .4505$\pm$2.2e-01 & No \\\cline{2-5}
&TensorFlow      &    -              &  -                & No  \\\hline
&Caffe-MPI       & .1019$\pm$2.4e-03 & .1325$\pm$2.7e-03 & Yes  \\\cline{2-5}
&CNTK            & .0485$\pm$9.9e-03 & .0595$\pm$1.9e-02 & No  \\\cline{2-5}
ResNet-50 &MXNet & .4994$\pm$1.1e-01 & .5561$\pm$3.7e-01 & No \\\cline{2-5}
&TensorFlow      &     -             &  -                & No  \\\hline
	\end{tabular}
		Notes: Due to \textit{grpc} hidden in TensorFlow, we could not get the accurate overhead of gredient communication accross multiple machines.
\end{threeparttable}
\end{table}
It is more challenging to hide the overhead of data communication across multiple servers since the bandwidth (or latency) of the network interface is much smaller (or longer) than PCIe or NVLink. In our experiments, the bandwidth of 56 Gbps InfiniBand is about a half of PCIe. Scaling performances across multiple machines are shown in Fig. \ref{fig:multilenodes}.

From Table \ref{table:mgtcomm}, it is noted that the communication time is hidden in Caffe-MPI, MXNet and TensorFlow. However, when scaling to multiple machines, the overhead of gradient aggregation across multiple machines could not be reduced easily, which is shown in Table \ref{table:mntcomm}. It is noted that the overhead of communication is not hidden in the inter-node environment except Caffe-MPI. Even though in the intra-node parallelism between gradient aggregation and backward propagation, the inter-node communication could cause the scaling performance drop down seriously. Both MXNet and TensorFlow use the PS method to synchronize the gradients across machines. The PS should collective the gradients from different machines via 56 Gbps InfiniBand, which has only 7 GB/s bandwidth and a high latency if the data transfer is not well optimized. Among the tested frameworks, they use different methodologies in communication across machines. Caffe-MPI implements the gradient aggregation with a decentralized method via efficient NCCL2.0, and it parallels with the backward propagation so that it can hide the communication. CNTK also uses NCCL2.0 to do the all-reduce, MXNet exploits TCP socket communication, and TensorFlow makes use of \textit{grpc}\footnote{grpc: https://grpc.io/} which is a high performance remote process call (RPC) framework. 

NCCL has high efficiency and low latency in doing collective communications via GPUDirect in the GPU cluster. For example, $t_{comm}$ of 2 GPUs and 4 GPUs on CNTK with AlexNet are 0.0906 and 0.236 respectively, and the size of gradients for communication is up to $S_g=63\times N_g$ MB. The all-reduce efficiency of CNTK (with NCCL2.0) is: 
\begin{equation}
E_{allreduce} \approx \cfrac{\frac{S_g}{t_{comm}}}{B_{rdma}}=\frac{S_g}{t_{comm}B_{rdma}}.
\end{equation}

It is known that $B_{rdma}=$56Gbps = 7GB/s, so $E_{allreduce}$ in 2 nodes (8 GPUs) and 4 nodes (16 GPUs) are $\frac{63\times 8}{0.0906\times 7\times 1024}=77.61\%$ and $\frac{63\times 16}{0.236\times 7\times 1024}=59.59\%$ respectively. However, the overhead of communication is also heavy compared to the time of computation, for example, $t_{comm}=0.1105$s and $t_{gpu}=0.0033+0.081+0.1780+0.0095=0.2722$s in GoogleNet. At last, the overall scaling efficiencies of CNTK are about 55\%, 67.5\% and 77.7\% in AlexNet, GoogleNet and ResNet-50 respectively when training on 4 machines.

MXNet exploits the customized KVStore \cite{li2014scaling} technique. Even though it makes the framework be equipped with the ability to scale to distributed environments easily, it also requires a very high quality of network to achieve better performance improvement or scalability. In the tested cases, when scaling to four machines, the communication overhead could become larger and leads to the low scaling efficiency due to the high latency and low actual bandwidth during the gradient communication. For example, the communication overhead is up to $0.4505$s, while the backward propagation only needs $0.1819$s in GoogleNet with four machines. The overhead of gradient aggregation cannot be hidden by backward propagation. Therefore, compared to its scalability of intra-node with multiple GPUs, MXNet performs lower scaling efficiency across multiple machines. As a result, MXNet achieves efficiencies of 35.625\%, 65.625\% and 35.6\% in AlexNet, GoogleNet and ResNet-50 respectively in our multi-node evaluation.

\textit{gprc} is the remote communication framework for TensorFlow, and RDMA used for TensorFlow may not be optimal, which results in relatively high latency compared to NCCL. Looking at the architecture of AlexNet and GoogleNet, the number of layers is small, and the computation of convolutional layers (big kernel size) is heavy. So it is easy to hide the latency of data copy in such scenarios. By contrast, ResNet-50 has deeper layers and smaller kernel sizes (most are $3\times 3$ and $1\times 1$ kernels) of convolutional layers, which requires more frequent communication of gradients but less computation of gradients, so the communication overhead is hard to hide since gradients of the previous layer are calculated too fast. Scaling to four machines, TensorFlow achieves scaling efficiencies of 50.625\%, 75.56\% and 52.187\% in AlexNet, GoogleNet and ResNet-50 respectively.

To summarize, not only the high-speed network is required to provide fast transfer of gradients, but it also gives a big challenge to the frameworks in optimizing the data communication across multiple machines to better utilize the hardware. Due to the high GFLOPS in a multi-GPU server (e.g., a server with 4 P40 GPUs), it makes the network and the implementation of S-SGD more challenging to achieve a high efficiency. 

\section{Conclusion and Future Work} \label{conclusionandfuturework}
In this work, we evaluate the performance of four popular distributed deep learning frameworks (Caffe-MPI, CNTK, MXNet and TensorFlow) over a 4-node dense GPU cluster (four Tesla P40 GPUs each node) connected with 56 Gbps InfiniBand via training three CNNs (AlexNet, GoogleNet and ResNet-50). We first build performance models to measure the speedup of synchronous SGD including different implementations from Caffe-MPI, CNTK, MXNet and TensorFlow. Then we benchmark the performances of these four frameworks covering single-GPU, multi-GPU and multi-machine environments. According to the experimental results and analysis, it shows some performance gaps among four different implementations, and there exist suboptimal methods that could be further optimized to improve the performance in evaluated frameworks in terms of I/O, cuDNN invoking, data communication across intra-node GPUs and inter-node GPUs.

For future work, we plan to evaluate the scalability of DL frameworks across low-bandwidth or high-latency networks (e.g., 1 Gbps Ethernet). And asynchronous SGD and model parallelism could also be considered.

\bibliographystyle{IEEEtran}
\Urlmuskip=0mu plus 1mu
\bibliography{dlbench-multigpu-datacom2018.bbl}
\end{document}